\documentclass[preprintnumbers,amsmath,11pt,amssymb,floatfix,superscriptaddress,nofootinbib]{revtex4}
\usepackage{graphicx}
\usepackage{epsfig}
\usepackage{bm}
\usepackage{amsfonts}

\begin{document}

\title{\Large New Modified Mimetic Gravity}

\author{Davood Momeni}
\email{d.momeni@yahoo.com}
\affiliation{Eurasian International Center for Theoretical Physics and Department of General
Theoretical Physics, Eurasian National University, Astana 010008, Kazakhstan.}

\author{Aziza Altaibayeva}
\email{aziza.bibol@mail.ru}
\affiliation{Eurasian International Center for Theoretical Physics and Department of General
Theoretical Physics, Eurasian National University, Astana 010008, Kazakhstan.}

\author{Ratbay Myrzakulov}
\email{rmyrzakulov@gmail.com}
\affiliation{Eurasian International Center for Theoretical Physics and Department of General
Theoretical Physics, Eurasian National University, Astana 010008, Kazakhstan.}

\begin{abstract}
Abstract:A modified Mimetic gravity (MMG) is proposed as a generalization of general relativity. The model contain a physical metric which is function of an auxiliary (unphysical)  metric and a Lyra's metric. We construct different kinds of conformally invariant models in different levels of the expansion parameter $\lambda$. This model phenomenologically  has been extended to higher order forms. Cosmology of a certain class of such models has been investigated in details. A cosmological solution has been proposed in inhomogenous form of scalar field. For homogenous case, energy conditions are widely investigated. We have shown that the system evaluated at intervals shorter than a certain time $T_c$ meets all the energy  conditions.  
\end{abstract}


\maketitle

%%%%%%%%%%%%%%%
\section{Introduction}
%%%%%%%%%%%%%%%%%
Modified gravity (see for recent reviews \cite{Nojiri:2010wj}
-\cite{Bamba:2014eea})
is the most popular way to adderess the recently observational data in favor of an accelerating expanding Universe \cite{obs1}-\cite{obs3} . It is believed that it is able to realize the acceleration expansion of the Universe  \cite{Nojiri:2008ku}-\cite{ijgmmp}. To describe dark energy and dark matter the first simple candidate is scalar field which it realizes modified gravity by reconstruction scheme. Scalar fields induce extra degree of freedom and cause ghost states which are physically unacceptable. Sometimes these new extra degrees of the freedom break the Lorentz symmetry in the Planck scales. So, they are able to improve the propagators of free graviton. In the absence of Lorentz symmetry , conformal symmetry is important and it is believed that \texttt{the small distance structure of canonical quantum gravity can be described using the conformal
group} \cite{Hoooft}. Not only in classical general relativity but in the Einstein-Cartan theories with torsion,it is possible to find a consistent Lagrangian of gravity as a gauge theory \cite{Maluf,Hehl,Poplawski,Momeni:2014taa}.\par
Recently a new motivated model for gravity has been introduced in which the geometry is Riemannian and it respects to the conformal symmetry as internal degree of freedom (not extra one) \cite{Chamseddine:2013kea}. The model is called as Mimetic gravity.  In this model it has been shown that the gravity can be described by a physical metric $g_{\mu\nu}$ in which the metric is function of the scalar field $\phi(x^{\alpha})$ (free and massless) and an auxiliarly metric $h_{\mu\nu}$. If we apply a generic conformal transformation on the auxiliary metric,it has been proven that the physical metric remains invariant. So, the theory is conformal invariant. By studying equation of motions for physical metric and scalar field we find two  equations in which the scalar field has unit norm of gradient i.e. $\partial_{\mu}\phi\partial^{\mu}\phi=1$. This constraint helps to the scalar field to be not an extra degree of freedom and consequently a non ghost field. In cosmological background we can show that dark matter appeares as an integration constant. In this sense it looks like a version of Horava-Lifshitz gravity \cite{Mukohyama:2009mz}. Different aspects of this model have been investigated in the literature 
\cite{Malaeb:2014vua}-\cite{Deruelle:2014zza}. Specially it was extended to the self- interacting scalar field \cite{Chamseddine:2014vna}.  Our aim in this paper is to generalize the Mimetic's idea to a more general case in which instead of scalar field we have another metric. This second metric is not neccesary to be Lorentzian like the physical and auxiliarly metric. We show that how it is posible to construct a generalized Mimetic gravity using a bi-metric approach in an arbitrary higher dimensional spacetime. The plan of this work is as the following: In Sec.II we construct generalized Mimetic gravity. Later we derive the equations of motion. In Sec.III we study cosmological solutions of a type of modified mimetic gravity. We show that the model has exact solutions for Hubble parameter and scalar field potential. In Sec. IV we investigate the validity of different types of energy conditions. We prove that there exist a range of time in which the system governs energy conditions. We summarize and conclude in Sec. V.

%%%%%%%%%%%%%%%%%%%%%%%
\section{Construction of  Mimetic model of gravity}
%%%%%%%%%%%%%%%%%%%%%%%
Let us to start by  the physical metric $g_{\mu\nu}$:
\begin{eqnarray}
g_{\mu\nu}=g_{\mu\nu}\Big(h_{\alpha\beta},\phi(x^{\alpha})\Big).
\end{eqnarray}
Here $h_{\alpha\beta}$ is an auxiliary metric and $\phi(x^{\alpha})\in\mathcal{R}$ is an uncharged scalar field. Here we assume that scalar field is frozen under conformal transformation. We suppose that under a conformal transformation , the auxiliarly metric transforms to the  $h_{\alpha\beta}\to \hat{h}_{\alpha\beta}=\Omega^2(x)h_{\alpha\beta},\ \ x\equiv x^a $,but the physical metric  has retained its original shape. It means ,
we are interested to have :
\begin{eqnarray}
g_{\mu\nu}\to\hat{g}_{\mu\nu}=g_{\mu\nu}.
\end{eqnarray}
One possible form of the metric is to be in the following form:
\begin{eqnarray}
g_{\mu\nu}=\sum_{n=1}^{\infty} c_n\Big[\lambda h^{\alpha\beta}Y_{\alpha\beta}\Big]^nh_{\mu\nu}.
\end{eqnarray}
In which $c_n$ are coefficients of a taylor series and $\lambda$ is a coupling constant under a conformal transformation as the following in $D$ dimensional spacetime we have:
\begin{eqnarray}
h_{\mu\nu}\longrightarrow\Omega^2(x)h_{\mu\nu} , (h_{\mu\nu})_{D\times D}.
\end{eqnarray}
We apply the conformal transformation to find:
\begin{eqnarray}
\hat{g}_{\mu\nu}=\sum_{n=1}^{\infty} c_n\Big[\lambda\Omega^{-2} h^{\alpha\beta}Y_{\alpha\beta}\Big]^n\Omega^2(x)h_{\mu\nu}=\sum_{n=1}^{\infty} c_n\Big[h^{\alpha\beta}Y_{\alpha\beta}\Big]^nh_{\mu\nu}\Omega^{2(1-n)}
\end{eqnarray}
here we study the form of auxiliarly metric $h_{\mu\nu},|h|=det(h_{\mu\nu})$. We know that:
\begin{eqnarray}
\hat{h}_{\mu\nu}=\Omega^2(x) h_{\mu\nu} \longrightarrow |\hat{h}|=\Omega^{2D}(x)|h_{\mu\nu}|=\Omega^{2D}(x)|h|
\end{eqnarray}
So we find:
\begin{eqnarray}
\Omega^2(x) =\Big(\frac{|\hat{h}|}{|h|}\Big)^{1/D} ,   \Omega^2(x) =\frac{\hat{h}_{\mu\nu}}{h_{\mu\nu}} \Longrightarrow \frac{\hat{h}_{\mu\nu}}{h_{\mu\nu}}\equiv\Big(\frac{|\hat{h}|}{|h|}\Big)^{1/D}
\end{eqnarray}
It is appropriate to write:
\begin{eqnarray}
\frac{\hat{h}_{\mu\nu}}{|\hat{h}|} = \frac{{h}_{\mu\nu}}{|h|} \longrightarrow \hat{h}_{\mu\nu}|\hat{h}|^{-1/D}= h_{\mu\nu}|h|^{-1/D}
\end{eqnarray}
Consequently we define a "new" auxiliary metric by: 
\begin{eqnarray}
H_{\mu\nu}= \frac{h_{\mu\nu}}{|h|^{1/D}}  \longrightarrow h_{\mu\nu} \equiv (h_{\mu\nu})_{D\times D}
\end{eqnarray}
Note that   
\begin{eqnarray}
H_{\mu\nu}\longrightarrow\hat{H}_{\mu\nu}=H_{\mu\nu}.
\end{eqnarray}
 because we know that:
\begin{eqnarray}
H^{\alpha\beta}\longrightarrow\Omega^{-2}(x)H^{\alpha\beta} 
,\ \ 
H_{\mu\nu}\longrightarrow\Omega^{-2}(x)H_{\mu\nu}.
\end{eqnarray}
thus the auxiliarly metric $H_{\mu\nu}$ is  self-conformal invariant. Now we rewite the physical metric in terms of the $H_{\mu\nu}$ as the following:
\begin{eqnarray}
g_{\mu\nu}=\sum_{n=1}^{\infty} c_n\Big[\lambda H^{\alpha\beta}Y_{\alpha\beta}\Big]^nH_{\mu\nu} \Longrightarrow (g)_{D\times D}=\Big(\sum_{n=1}^{\infty} c_n\Big[\lambda H^{\alpha\beta}Y_{\alpha\beta}\Big]^n\Big)(H)_{D\times D} .
\end{eqnarray}
To have a closed form for metric we choice $c_n=\frac{1}{n!}$. So we obtain:
\begin{eqnarray}
g_{\mu\nu}=exp\Big[\lambda H^{\alpha\beta}Y_{\alpha\beta}\Big]H_{\mu\nu}\label{g}.
\end{eqnarray}
We presented  the most general form of metrics which remain invariant under conformal transformations. We mention here that it is possible to  calculate the coefficents of the series by derivatives with respect to the $\lambda$. For example we have:
\begin{eqnarray}
\frac{\delta g_{\mu\nu}}{\delta\lambda}=(H^{\alpha\beta}Y_{\alpha\beta}) exp\Big[\lambda H^{\alpha\beta}Y_{\alpha\beta}\Big]H_{\mu\nu}\Big|_{\lambda=0}=(H^{\alpha\beta}Y_{\alpha\beta})H_{\mu\nu}.
\end{eqnarray}
For second order we obtain:
\begin{eqnarray}
\frac{\delta^2 g_{\mu\nu}}{\delta\lambda^2}\Big|_{\lambda=0}=(H^{\alpha\beta}Y_{\alpha\beta})^2H_{\mu\nu}.
\end{eqnarray}
Where $(H^{\alpha\beta}Y_{\alpha\beta})^2H_{\mu\nu}$ defines the second order Mimetic model. We mention here that:
\begin{eqnarray}
\frac{\delta \hat{g}_{\mu\nu}}{\delta\lambda}\Big|_{\lambda=0}=\frac{\delta g_{\mu\nu}}{\delta\lambda}\Big|_{\lambda=0}
\end{eqnarray}
It is possible to write the physical metric in the following series form:
\begin{eqnarray}
g_{\mu\nu}=H_{\mu\nu}+\lambda(H^{\alpha\beta}Y_{\alpha\beta})H_{\mu\nu}+\frac{\lambda^2}{2!}(H^{\alpha\beta}Y_{\alpha\beta})^2H_{\mu\nu} +...\label{series}.
\end{eqnarray}
Here the first term is invariant under conformal transformation. The second term is also invariant under such transformations thanks to the self invariance of the auxiliary mrtric $\hat{H}_{\mu\nu}=H_{\mu\nu} $. We sumarize the results:\par
We constructed a model for gravity defined by a physical metric $g_{\mu\nu}$ which it remains invariant under consormal transformations of an auxiliary metric as the following:
\begin{eqnarray}
H_{\alpha\beta}\longrightarrow\hat{H}_{\alpha\beta}=H_{\mu\nu} \Longleftrightarrow \hat{g}_{\mu\nu} = g_{\mu\nu} \Longleftrightarrow H_{\mu\nu}=\frac{h_{\mu\nu}}{|h|^{1/D}}
\end{eqnarray}
It is interesting to mention here that if we choice the metric $Y_{\alpha\beta}=\partial_{\alpha}\phi\partial_{\beta}\phi$ (Lyra's geometry \cite{Lyra}) then the series up to the first order of $\mathcal{O}(\lambda)$ is the one proposed by \cite{Chamseddine:2013kea}. But the second term is a higher order non-trivial extension of the original Mimetic model. So, we can say:\par
\texttt{Mimetic gravity is a bi-metric model (with two auxiliary metric) ,one is Lorentzian and one is Lyra's metric}.

%%%%%%%%%%%%%%%%%%%%%%%
\subsection{Equations of the motion}
%%%%%%%%%%%%%%%%%%%
In this section we derive the gravitational field equations for a pair of auxiliary metric and physical metric. We start by the common form of Einstein-Hilbert action in $D\geq4$ dimensins with a matter Lagrangian $\mathcal{L}_m$ in  the following form:
\begin{eqnarray}
S=\frac{1}{2k^2} \int \Big\{R\Big[g(H_{\alpha\beta},Y_{\alpha\beta})\Big]+\mathcal{L}_m \Big\}\sqrt{-g(H_{\alpha\beta},Y_{\alpha\beta})}d^Dx\label{S}.
\end{eqnarray}
By varying the (\ref{S}) with respect to the physical metric $g_{\alpha\beta}$ we find:
\begin{eqnarray}
\delta S=\frac{1}{2k^2} \int d^4x\sqrt{-g(H_{\alpha\beta},Y_{\alpha\beta})}\Big(G^{\alpha\beta}-T^{\alpha\beta})\Big) \delta g_{\alpha\beta}(H_{\mu\nu},Y_{\mu\nu})
\end{eqnarray}
Where $T^{\alpha\beta}=-\frac{2}{\sqrt{-g}}\frac{\delta S}{\delta g_{\alpha\beta}}$ denotes the typical energy momentum tensor of matter fields.  But since $g=g(H,Y)$, so it is needed to compute $\delta g_{\alpha\beta}$. Using (\ref{g}) we find:
\begin{eqnarray}
&&\delta g_{\alpha\beta}=exp(H^{\alpha\beta}Y_{\alpha\beta})\delta H_{\mu\nu} +(\delta H^{\alpha\beta})Y_{\alpha\beta}exp(H^{\alpha\beta}Y_{\alpha\beta})H_{\mu\nu}
\\&&\nonumber+H^{\alpha\beta}\delta Y_{\alpha\beta}exp(H^{\alpha\beta}Y_{\alpha\beta})H_{\mu\nu}
=exp(H^{\alpha\beta}Y_{\alpha\beta})\Big[\delta H_{\mu\nu}+(Y_{\alpha\beta}\delta H^{\alpha\beta})H_{\mu\nu}+H^{\alpha\beta}H_{\mu\nu}\delta Y_{\alpha\beta}\Big] 
\end{eqnarray}
We know that:
\begin{eqnarray}
\delta H^{\alpha\beta}=g^{\acute{\mu}\alpha}g^{\acute{\nu}\beta}\delta H_{\acute{\mu}\acute{\nu}}\longrightarrow  Y_{\alpha\beta}\delta H^{\alpha\beta}=Y^{\acute{\mu}\acute{\nu}}\delta H_{\acute{\mu}\acute{\nu}}
\end{eqnarray}
So, in fact we have:
\begin{eqnarray}
\delta H_{\mu\nu}=\delta_\mu^\alpha\delta_\nu^\beta \delta H_{\alpha\beta}
\end{eqnarray}
Using these identities we find the total variation of (\ref{S}) as the following:
\begin{eqnarray}
&&\delta S=\frac{1}{2k^2} \int d^4x\sqrt{-g}(G^{\alpha\beta}-T^{\alpha\beta}) exp(\lambda H^{\alpha\beta}Y_{\alpha\beta})_\lambda \Big(\delta H_{\mu\nu}+\\&&\nonumber+(Y_{\alpha\beta}g^{\mu\alpha}g^{\nu\beta}\delta H_{\mu\nu})H_{\mu\nu}+H^{\alpha\beta}H_{\mu\nu}\delta Y_{\alpha\beta}\Big)=0
\end{eqnarray}
Or equivalently we write it in the following form:
\begin{eqnarray}
\delta S=\frac{1}{2k^2} \int d^4x\sqrt{-g_D}\lambda(G^{\alpha\beta}-T^{\alpha\beta}) exp(\lambda H^{\alpha\beta}Y_{\alpha\beta})\Big[\delta H_{\mu\nu}+\notag\\+(Y^{\alpha\beta}\delta H_{\alpha\beta})H_{\mu\nu}+H^{\alpha\beta}H_{\mu\nu}\delta Y_{\alpha\beta}\Big]=0
\end{eqnarray}
Note that:
\begin{eqnarray}
\Big[\delta H_{\mu\nu}+(Y^{\alpha\beta}\delta H_{\alpha\beta})H_{\mu\nu}+H^{\alpha\beta}H_{\mu\nu}\delta Y_{\alpha\beta}\Big]=\notag\\=\delta_\mu^\alpha\delta_\nu^\beta \delta H_{\alpha\beta}+ Y^{\alpha\beta}\delta H_{\alpha\beta}H_{\mu\nu}+H^{\alpha\beta}H_{\mu\nu} \delta Y_{\alpha\beta}
\end{eqnarray}
So we obtain:
\begin{eqnarray}
\Big[\delta H_{\mu\nu}+(Y^{\alpha\beta}\delta H_{\alpha\beta})H_{\mu\nu}+H^{\alpha\beta}H_{\mu\nu}\delta Y_{\alpha\beta}\Big]=\notag\\=(\delta_\mu^\alpha\delta_\nu^\beta + Y^{\alpha\beta}H_{\mu\nu})\delta H_{\alpha\beta}+H^{\alpha\beta}H_{\mu\nu}\delta Y_{\alpha\beta}
\end{eqnarray}
Finally we have:
\begin{eqnarray}
\delta g_{\mu\nu}=\Big[(\delta_\mu^\alpha\delta_\nu^\beta + Y^{\alpha\beta}H_{\mu\nu})\delta H_{\alpha\beta}+H^{\alpha\beta}H_{\mu\nu}\delta Y_{\alpha\beta}\Big]exp(\lambda H^{\alpha\beta}Y_{\alpha\beta})
\end{eqnarray}
And consequently we obtain:
\begin{eqnarray}
&&\delta S=\frac{1}{2k^2} \int d^4x\sqrt{-g_D}exp(\lambda H^{\alpha\beta}Y_{\alpha\beta})\Big[(G^{\mu \nu}-T^{\mu \nu}) (\delta_{\mu}^{\acute{\alpha}} \delta_{\nu}^{\acute{\beta}} +Y^{\acute{\alpha}\acute{\beta}}H_{\mu\nu})\delta H_{\acute{\alpha}\acute{\beta}}+	\\&&\nonumber+H^{\acute{\alpha}\acute{\beta}}H_{\mu\nu}\delta Y_{\acute{\alpha}\acute{\beta}}(G^{\mu\nu}-T^{\mu\nu})\Big]
	\\&&\nonumber\Rightarrow
\delta S=\frac{1}{2k^2} \int d^4x\sqrt{-g_D}exp[\lambda H^{\alpha\beta}Y_{\alpha\beta}]\Big[(G^{\acute{\alpha}\acute{\beta}}-T^{\acute{\alpha}\acute{\beta}})+Y^{\acute{\alpha}\acute{\beta}}T^{\mu \nu}H_{\mu\nu})\delta H_{\acute{\alpha}\acute{\beta}}+	\\&&\nonumber+H^{\acute{\alpha}\acute{\beta}}\delta Y_{\acute{\alpha}\acute{\beta}}H_{\mu\nu}(G^{\mu\nu}-T^{\mu\nu})\Big]
\end{eqnarray}
So, we find the following variations of (\ref{S}) for $\{Y,g\}$:
\begin{eqnarray}
&&\delta S_{Y}\propto\int d^4x\sqrt{-g_D}exp[\lambda H^{\alpha\beta}Y_{\alpha\beta}]H^{\acute{\alpha}\acute{\beta}}H_{\mu\nu}(G^{\mu\nu}-T^{\mu\nu})\delta Y_{\acute{\alpha}\acute{\beta}}	\\&&\nonumber
\delta S_g\propto\int d^4x\sqrt{-g_D}exp[\lambda H^{\alpha\beta}Y_{\alpha\beta}]\delta H_{\acute{\alpha}\acute{\beta}}\Big[(G^{\acute{\alpha}\acute{\beta}}-T^{\acute{\alpha}\acute{\beta}})+Y^{\acute{\alpha}\acute{\beta}}T^{\mu \nu}H_{\mu\nu}\Big]
\end{eqnarray}
Putting zero we derive these quantities:
\begin{eqnarray}
&&\frac{\delta S_{H} }{\delta H_{\acute{\alpha}\acute{\beta}}}=0\longrightarrow(G^{\acute{\alpha}\acute{\beta}}-T^{\acute{\alpha}\acute{\beta}})+Y^{\acute{\alpha}\acute{\beta}}T^{\mu \nu}H_{\mu\nu}=0\label{eom1}
\\&&
\frac{\delta S_{Y} }{\delta Y_{\acute{\alpha}\acute{\beta}}}=0\longrightarrow H^{\acute{\alpha}\acute{\beta}}H_{\mu\nu}(G^{\mu\nu}-T^{\mu\nu})=0\label{eom2}.
\end{eqnarray}
Trace of (\ref{eom1}) gives us:
\begin{eqnarray}
(G-T)+ Y_{\acute{\alpha}}^{\acute{\alpha}} T^{\mu\nu}H_{\mu\nu}=0
\end{eqnarray}
But (\ref{eom2}) reads as an orthogonality condition. It is adequate to write (\ref{eom2}) as the following:
\begin{eqnarray}
&&H_{\alpha}^{\alpha}H_{\mu\nu}G^{\mu\nu}-H_{\alpha}^{\alpha}H_{\mu\nu}T^{\mu\nu}=0\longrightarrow
H_{\alpha}^{\alpha}\neq0 \longrightarrow H_{\mu\nu}G^{\mu\nu}=H_{\mu\nu}T^{\mu\nu}
\end{eqnarray}
Equations (\ref{eom1},\ref{eom2}) complete the physical description of the model is defined by (\ref{S}).

%%%%%%%%%%%%%%%%%%%%%%%%%5%%%%%%%%%
\section{Modified Mimetic Cosmology in flat spacetime}
%%%%%%%%%%%%%%%%%%%%%%%%%%%%%%%%%%%
This section is devoted to cosmological solution of a type a MMG ,proposed in \cite{Chamseddine:2014vna}. 
Let us to start by  the FRW metric in the following form:
\begin{eqnarray}
g_{\mu\nu}=diag\Big(1,-a^2(t)\Sigma_3\Big).
\end{eqnarray}
Consider the theory with the following form,is called as MMG,
\begin{equation}
S=%
%TCIMACRO{\dint }%
%BeginExpansion
{\displaystyle\int}
%EndExpansion
d^{4}x\sqrt{-g}\left[  -\frac{1}{2}R\left(  g_{\mu\nu}\right)  +\lambda\left(
g^{\mu\nu}\partial_{\mu}\phi\partial_{\nu}\phi-1\right)  -V\left(
\phi\right)  +\mathcal{L}_{m}\left(  g_{\mu\nu},...\right)  +\frac{1}{2}\gamma\left(  \square\phi\right)  ^{2}\right]  \ ,
\end{equation}
where $\gamma$ is a constant. 
Following \cite{Chamseddine:2014vna},the $0-0$ , $0-i,\ \ (i=x,y,z)$ and $i-i$ components of the Einstein equation read:
\begin{eqnarray}
0-0:,\ \ && 3H^2=V(\phi)-\gamma(1\pm\frac{k}{a})\frac{d}{dt}(H(3\pm\frac{2k}{a}))+2\lambda(1\pm\frac{k}{a})^2+\frac{\gamma H^2}{2}(3\pm \frac{2k}{a})^2,\\
0-i:,\ \ && 2\lambda\dot{\phi}\acute{\phi}-\gamma\acute{\phi}\dot{\chi}=0,\\
i-i:,\ \ &&2\dot{H}+3H^2=\frac{2\lambda(k_x)^2}{a^2}-V(\phi)\\&&\nonumber-\gamma\Big[(1\pm\frac{k}{a})\Big((3\pm\frac{2k}{a})\dot{H}\mp\frac{2k}{a}H^2\Big)+\frac{1}{2}H^2(3\pm\frac{2k}{a})^2\Big].
\end{eqnarray}
Here $\chi=\Box\phi=(-g)^{-1/2}\partial_{\mu}((-g)^{1/2}\partial^{\mu}\phi)$.
Using the $(0-i)$ equation we find:
\begin{eqnarray}
\ \ &&\acute{\phi}(2\lambda\dot{\phi}-\dot{\chi})=0,\\
\ \ &&\acute{\phi}=k_x\neq0 \longrightarrow 2\lambda\dot{\phi}-\dot{\chi}=0.
\end{eqnarray}
The scalar field constraint reads:
\begin{eqnarray}
&&g^{\mu\nu}\partial_{\mu}\partial_{\nu}=1,\ \
(\frac{\partial\phi}{\partial t})^2-\frac{1}{a^{2}(t)}(\nabla\phi)^2=1.
\end{eqnarray}
Where $\nabla\phi=(\partial_x\phi,\partial_y\phi,\partial_z\phi)$. We can take $\phi=\phi(t;x^a),\ \ x^a=\{x,y,z\}$, so the exact solution for the normalized scalar field reads:
\begin{eqnarray}
\phi(t,\vec{x})=t+\vec{k}\cdot\vec{x}\pm|\vec{k}|\int{\frac{dt}{a(t)}}\label{phi}.
\end{eqnarray}
Using (\ref{phi}) and $(0-i)$ equation we obtain:
\begin{eqnarray}
 &&2\lambda\phi(t,\vec{x})- \chi(t)=f(x) =\texttt{Arbitrary funcsion}
\end{eqnarray}
Now by plugging (\ref{phi}) in the above equation we obtain: 
\begin{eqnarray}
 &&2\lambda (t\pm k\int\frac{dt}{a(t)}+\vec{k}\cdot\vec{x})-\gamma H(3\pm\frac{2k}{a})=f(x),	\Rightarrow   f(x) = 2\lambda \vec{k}\cdot\vec{x},\\
\ \ &&\Rightarrow t\pm k\int\frac{dt}{a(t)}=\frac{\gamma}{2\lambda}(3\pm\frac{2k}{a})H,
\end{eqnarray}
Using the another FRW equation we obtain:
\begin{eqnarray}
&& 1\pm \frac{k}{a}=\frac{\gamma}{2\lambda}\Big[3\pm\frac{2k}{a}\Big]\dot{H}+\frac{\gamma}{2\lambda}(\mp 2k\frac{H^2}{a}),
\end{eqnarray}
Finally it is possible to write the FRW equation using the effective pressure and energy density as the following:
\begin{eqnarray}
&& 3H^2=\kappa^2\rho_{eff},\ \ 2\dot{H}=-\kappa^2(p_{eff}+\rho_{eff}).
\end{eqnarray}
Where the effective quantities are written as the following:
\begin{eqnarray}
p_{eff}=V(\phi)\Big[\frac{3\pm\frac{2k}{2}}{\frac{\gamma}{2}(1\pm \frac{k}{a})+\frac{1}{3\pm\frac{2k}{2}}-\frac{3}{4k}\pm\frac{\gamma}{2}-\frac{a\gamma}{8k}(3\pm\frac{2k}{2})^2}\Big]-\notag\\-\lambda\Big[\frac{\frac{9a^2}{4k^2\gamma}(1\pm \frac{k}{a})+\frac{2k_x^2}{a^2(3\pm\frac{2k}{2})}}{\frac{\gamma}{2}(1\pm \frac{k}{a})+\frac{1}{3\pm\frac{2k}{2}}-\frac{3}{4k}\pm\frac{\gamma}{2}-\frac{a\gamma}{8k}(3\pm\frac{2k}{2})^2}\Big]
\end{eqnarray}
\begin{eqnarray}
\rho_{eff}=V(\phi)\Big[\frac{3a(3\pm\frac{2k}{2})^2}{2k\gamma(1\pm \frac{k}{a})+\frac{4k}{3\pm\frac{2k}{2}} -3a \pm 2k\gamma-\frac{a\gamma}{2}(3\pm\frac{2k}{2})^2}\Big]-\notag\\-\frac{3a\lambda}{k}\Big[\frac{(1\pm \frac{k}{a})}{4\gamma}-\frac{\frac{9a^2}{4k^2\gamma}(1\pm \frac{k}{a})+\frac{2k_x^2}{a^2(3\pm\frac{2k}{2})}}{2\gamma(1\pm \frac{k}{a})+\frac{4}{3\pm\frac{2k}{2}} -\frac{3a}{k} \pm 2\gamma-\frac{a\gamma}{2k}(3\pm\frac{2k}{2})^2}\Big]
\end{eqnarray}
Specially when $k=0$ we have:
\begin{eqnarray}
p_{eff}=2V(\phi)+2\lambda,\ \
\rho_{eff}=\frac{2+6\gamma}{2+3\gamma}V(\phi)+2\lambda
\end{eqnarray}
In this case the exact solution for FRW equations are:
\begin{eqnarray}
 \dot{H}=\frac{2\lambda}{3\gamma}\Longrightarrow H(t)=\frac{2\lambda}{3\gamma}(t-t_0)\Longrightarrow a(t)=exp\Big(\frac{\lambda}{3\gamma}(t-t_0)^2\Big)	\label{k0-eq2}.
\end{eqnarray}
This equation has some solution for a(t). If 
we substitue $\dot{H}=\frac{2\lambda}{3\gamma}$ in $0-0$ ($k=0$) we find:
\begin{eqnarray}
&&3H^2-\frac{9}{2}\gamma H^2=V(t)\Longrightarrow
V(t)=3(1-\frac{3}{2}\gamma)H^2=(1-\frac{3}{2}\gamma)(\frac{2\lambda}{3\gamma})^2(t-t_0)^2.	 
\end{eqnarray}
If  $\phi=t$ we find:
\begin{eqnarray}
 && V(\phi)=\frac{4\lambda^2}{3\gamma^2}(1-\frac{3\gamma}{2})(\phi-\phi_0)^2	
\end{eqnarray}
So MMG posses only the scalar potential in the form of square $(\phi-\phi_0)^2	$. But for  $k\neq0$  the bahavior is different,we've:
\begin{eqnarray}
0-i:,\ \ && \frac{\gamma}{2\lambda}(3\pm\frac{2k}{a})\dot{H}\mp \frac{k}{a}(1+\frac{\gamma H^2}{\lambda})-1=0,
\end{eqnarray}
The equation can be solved to obtain H. There are two branches $\pm$ and consequently the pair of solutions for $H_{\pm}(t)$ as the following:

\begin{eqnarray}
\pm\sqrt{\gamma}\int_{a_0}^{a_{+}(t)}{\frac{3x+2k}{x\sqrt{c_1\gamma x^2-4\lambda k^2+12\lambda x^2\log x-20 k\lambda x}}}-t-c_2=0\label{a1}\\
\pm\sqrt{\gamma}\int_{a_0}^{a_{-}(t)}{\frac{3x-2k}{x\sqrt{c'_1\gamma x^2-4\lambda k^2+12\lambda x^2\log x+20 k\lambda x}}}-t-c'_2=0\label{a2}.
\end{eqnarray}
Here $\{c_i,c'_i\},\ \ i=1,2$ are constants. Using (\ref{a1},\ref{a2}) we are able to find $\{H_{+},H_{-}\}$. We find:
\begin{eqnarray}
H_{+}(t)=\pm\sqrt{\gamma}^{-1}(3a(t)+2k)^{-1}\sqrt{c_1\gamma a(t)^2-4\lambda k^2+12\lambda a(t)^2\log a(t)-20 k\lambda a(t)}\label{H+}\\
H_{-}(t)=\pm\sqrt{\gamma}^{-1}(3a(t)-2k)^{-1}\sqrt{c'_1\gamma a(t)^2-4\lambda k^2+12\lambda a(t)^2\log a(t)+20 k\lambda a(t)}\label{H-}
\end{eqnarray}

%%%%%%%%%%%%%%%%%%%%%%%%%%%%%%%%%%%%%%%%%%%%%%%%%%%%%%%%%%%%%%%%
\section{Energy conditions in modified  mimetic cosmology}
%%%%%%%%%%%%%%%%%%%%%%%%%%%%%%%%%%%%%%%%%%%%%%%%%%%%%%%%%%%%%%%%%55
Consider a  type of modified gravity in cosmological background with the following forms of effective FRW equations:
\begin{eqnarray}
3H^2=\kappa^2\rho_{eff},\ \ 2\dot{H}=-\kappa^2(p_{eff}+\rho_{eff}).
\end{eqnarray}
Here the effective quantities are functions of the geometrical quantities like $\{H,\dot{H},...,H^{(n)},a(t),..\}$. 

The null energy condition
(NEC), weak energy condition (WEC), strong energy condition (SEC) and the dominant energy
condition (DEC) are given by \cite{lobo,anzhong}:
\begin{eqnarray}
\text{NEC}&\Longleftrightarrow&\rho_{\text{eff}}+p_{\text{eff}}\geq0.\label{n1}\\
\text{WEC}&\Longleftrightarrow& \rho_{\text{eff}}\geq0\ \text{and}\ \rho_{\text{eff}}+p_{\text{eff}}\geq0.\label{n2}\\
\text{SEC}&\Longleftrightarrow& \rho_{\text{eff}}+3p_{\text{eff}}\geq0\ \text{and}\ \rho_{\text{eff}}+p_{\text{eff}}\geq0.\label{n3}\\
\text{DEC}&\Longleftrightarrow& \rho_{\text{eff}}\geq0\ \text{and}\ \rho_{\text{eff}}\pm p_{\text{eff}}\geq0.\label{n4}
\end{eqnarray}
It is clear to us that these energy conditions are independent from any gravity theory and that these are purely geometrical \cite{hawking}. Furthermore, NEC implies WEC and WEC implies SEC and DEC. In  MMG we clearly suppose that  the non exotic ordinary matter  satisfies all the energy conditions separately i.e. $\rho_m\geq0$, $\rho_m\pm p_m\geq0$, $\rho_m+3p_m\geq0$. Different types of modified gravities have been checked for such energy conditions, namely $f(R)$ theory  \cite{wang1}, $f(T)$ in which $T$ denotes torsion 
\cite{Jamil:2012ck} and more(see \cite{Nojiri:2006ri} for a review).  One  checks standard energy conditions
on Quantum de Sitter cosmology and phantom matter and energy conditions in phantom / tachyon inflationary cosmology perturbed by quantum effects\cite{hep-th/0303117,hep-th/0306212}. Furthermore stronger analysing including non-linear energy conditions has been investigated \cite{Martin-Moruno:2013sfa}. 
In our work we also would like to check these conditions. \par
Let us to think on a case with $\vec{k}=0$. So,we write:
\begin{eqnarray}
\phi(t,\vec{x})=t.
\end{eqnarray}
The effective quantities read :
\begin{eqnarray}
\ \ && p_{eff} = -2\lambda-m^2t^2\\
\ \ &&\rho_{eff}=\frac{(4-2\gamma)\lambda+(1-3\gamma)m^2t^2}{2-3\gamma} 
\end{eqnarray}
The energy conditions read:
\begin{eqnarray}
 && NEC \Longleftrightarrow \frac{4\gamma\lambda-m^2t^2}{2-3\gamma} \geq0.\\
\ \ && WEC \Longleftrightarrow \frac{(4-2\gamma)\lambda+(1-3\gamma)m^2t^2}{2-3\gamma}\geq0    \  \   \&    \  \     \frac{4\gamma\lambda-m^2t^2}{2-3\gamma}\geq0.\\
\ \ && SEC \Longleftrightarrow \frac{(-8+16\gamma)\lambda+(-5+6\gamma)m^2t^2}{2-3\gamma}\geq0 \  \   \&    \  \  \frac{4\gamma\lambda-m^2t^2}{2-3\gamma}\geq0.\\
\ \ && DEC \Longleftrightarrow \frac{(4-2\gamma)\lambda+(1-3\gamma)m^2t^2}{2-3\gamma}\geq0   \  \   \&    \  \ (\frac{4-2\gamma}{2-3\gamma}\pm2)\lambda+(\frac{1-3\gamma}{2-3\gamma}\pm1)m^2t^2 \geq0.\\
\end{eqnarray}
In the case of $2-3\gamma>0$ we conclude that all energy conditions satisfy for a time scale no longer than $t_c$,where
\begin{eqnarray}
t_c\leq\frac{1}{m}\sqrt{4\gamma\lambda}.
\end{eqnarray}
\par
Now we consider the case of $\gamma> 2/3$,so we obtain:
\begin{eqnarray}
\ \ && NEC \Longleftrightarrow \texttt{is satisfied}.\\
\ \ && WEC \Longleftrightarrow  \texttt{is satisfied when}\     \    t\leq t^{*}_1,\   \      t^{*}_1=\frac{1}{m}\sqrt{\frac{4-2\gamma}{1-3\gamma}\lambda}.\\
\ \ && SEC \Longleftrightarrow \texttt{is satisfied when} \      \       t\leq t^{*}_2,\   \      t^{*}_2=\frac{2}{m}\sqrt{\frac{16\gamma -12}{6\gamma -5}\lambda}.\\
\ \ && DEC \Longleftrightarrow \texttt{is satisfied when}\         \        t\leq t^{*}_{3},\   \      t^{*}_3=\frac{1}{m}\sqrt{\frac{8-8\gamma}{3-6\gamma}\lambda}\}.
\end{eqnarray}
The NEC does not violate. It is a posibility to preserve the rest,when we have:
\begin{eqnarray}
t\leq \texttt{Max}\{  t^{*}_1,  t^{*}_2,t^{*}_{3}\}.
\end{eqnarray}
So, we should find the following quantity in fixed $\{m,\lambda>0\}$:
\begin{eqnarray}
T_c=\texttt{Max}\{ \sqrt{\frac{16(4\gamma -3)}{6\gamma -5}},\sqrt{\frac{2(2-\gamma)}{1-3\gamma}} ,\sqrt{\frac{8(1-\gamma)}{3(1-2\gamma)}}\}_{\gamma> 2/3}=\sqrt{\frac{2(2-\gamma)}{1-3\gamma}}.
\end{eqnarray}
So, for time scales shorter than $T_c$ all kinds of energy conditions are  satisfied.

%%%%%%%%%%%%%%%%%%%%%%%%%%%%%%%%%%%%%%%%%%%%%%%%%%%%%%%%%%%%%%%%%%%%
\section{Final remarks}
%%%%%%%%%%%%%%%%%%%%%%%%%%%%%%%%%%%%%%%%%%%%%%%%%%%%%%%%%%%%%%%%%%%%%%%
There are different types of modified gravity theories like $f(R),f(T),f(R,G)$ and more. In all these theories there exists one or more extra degrees of freedom. Only a few numbers of such models are ghost free. Specially the case of a single or multiple scalar fields are considered as potentially interesting models for dark sectors like  energy or matter. In this work we have been motivated by a very recently proposal of gravity as a mimetic model with conformal symmetry (internal symmetry). We proposed a generalized model for gravity at large scales. In our proposal, the metric is a function of an auxiliay metric and another Lyra's metric. When we apply the conformal symmetry on the auxiliary metric we find that the physical metric is invariant. The gravitational action is obtained by replacing the Ricci scalar of the physical metric. But to find the equations of motion we should take into the account the total variation of the metric. We find two independent equations of motion for auxiliary and Lyra metrics. If we specify Lyra's metric as the bi -linear combination of the gradient of a scalar field like $Y_{\alpha\beta}=\partial_{\alpha}\phi\partial_{\beta}\phi$ we find that the scalar field must have a unit norm. So,  with a fixed physical metric, the scalar metric has a definite form. There is no additional second order differential equation for the scalar field to specify its dymanics. Further more,we investigate some particular solutions of a type of Mimetic gravity for cosmology. We show that with a homogenous scalar field the scalar field interaction is a square function of $\phi$. In this case the Hubble parameter is linear as a function of time. For a slightly inhomogenous scalar field,we have a couple forms of Hubble parameter. The form of scalar potential is determined from FRW equations.  Later we investigate the validity of energy conditions in this kind of modified gravitiy. We show that it is possible to rewrite the pair of FRW equations in an effective form. By using the effective energy density and pressure we study energy conditions. We showed that for time scales shorter than a certain time the system respects all energy conditions. For longer times these energy conditions are violated. It predicts that in the gravitational system there is a phase transition in cosmological scales.

%%%%%%%%%%%%%%%%%%%%%%%%%%%%%%
\subsection{Acknowledgement}
D. Momeni
would like to thank the kind hospitality of the Abdus Salam ICTP, Trieste, Italy where part
of this work was completed. We would like to thank  R. da Rocha, S. D. Odintsov for useful comments and also the anonymous reviewer for enlightening comments related to this work. 
%%%%%%%%%%%%%%%%%%%%%%%%%%%%%%%%

\end{document}